\documentclass[preprint2,6pt]{emulateapj}%
\usepackage{graphicx}
\usepackage{amsmath}
\usepackage{amsfonts}
\usepackage{amssymb}%
\providecommand{\U}[1]{\protect\rule{.1in}{.1in}}

\newcommand{\beq}{\begin{equation}}
\newcommand{\eeq}{\end{equation}}
\newcommand{\ba}{\begin{array}}
\newcommand{\ea}{\end{array}}

\newcommand{\na}{New~Astr.}

\begin{document}

\title{Constraining Magnetization of Gamma-Ray Bursts Outflows using Prompt Emission Fluence}
\shorttitle{}
\author{Asaf Pe'er \altaffilmark{1}  }
\shortauthors{Pe\'er}

\altaffiltext{1}{Physics Department, University College Cork, Cork, Ireland}

\begin{abstract}
I consider here acceleration and heating of relativistic outflow by
local magnetic energy dissipation process in Poynting flux dominated
outflow. Adopting the standard assumption that the reconnection rate
scales with the Alfv\'en speed, I show here that the fraction of
energy dissipated as thermal photons cannot exceed $(13 \hat \gamma
-14)^{-1} = 30\%$ (for adiabatic index $\hat \gamma = 4/3$) of the
kinetic energy at the photosphere. Even in the most radiatively
efficient scenario, the energy released as non-thermal photons during
the prompt phase is at most equal to the kinetic energy of the
outflow. These results imply that calorimetry of the kinetic energy
that can be done during the afterglow phase, could be used to
constrain the magnetization of gamma-ray bursts (GRB) outflows. I
  discuss the recent observational status, and its implications on
  constraining the magnetization in GRB outflows.
\end{abstract}
\maketitle

\section{Introduction}

One of the key open questions in the study of relativistic outflows is
the mechanism responsible for accelerating the plasma to
ultra-relativistic speeds, with inferred Lorentz factor $\Gamma
\gtrsim$~few tens in active galactic nuclei (AGNs), and $\Gamma
\gtrsim 100$ in gamma-ray bursts (GRBs). In the classical GRB
``fireball'' model, for example, the outflow is accelerated by
radiative pressure, and magnetic fields are sub-dominant \citep[for
  several recent reviews see][and references therein]{ MR14, KZ15,
  Peer15}. On the other hand, in recent years models in which GRB
outflows are Poynting flux dominated became increasingly popular
\citep{Levinson06a, Lyutikov06, Giannios08, TMN08,
  Komissarov+09, Metzger+11, ZY11, MU12, Sironi+15}.

There are indeed strong theoretical arguments in favor of Poynting
flux dominated flows in GRBs. First, it is well established that the
progenitor of a GRB is a compact object of solar scale, namely a black
hole or neutron star. A Poynting flux dominated outflow will naturally
occur if the compact object rotates and possesses a magnetic field
\citep{BZ77, BP82}. The jet could tap into the rotational energy of
the neutron star, black hole or accretion disk through the agency of
an ordered magnetic field that threads the source
\citep[e.g.,][]{Usov92, Thompson94, VK01, Drenkhahn02, DS02, LB03,
  VK03}. Second, a well known problem of non-magnetized outflow models
is the very low efficiency in converting the kinetic energy to the
observed radiation. This must follow an episode(s) of kinetic energy
dissipation. However, the leading dissipation mechanism, namely
internal shock waves \citep{RM94} are known to be inefficient -
typically, only a few \% of the kinetic energy is dissipated
\citep{MMM95, Kobayashi+97, PSM99}. In Poynting flux dominated flows
on the other hand, dissipation of magnetic energy can take place via a
reconnection process. This process is known to provide an efficient
way of converting the energy stored in the magnetic field
\citep{Drenkhahn02, DS02, Komissarov+09, Lyubarsky10a, TNM10b,
  MU12, Sironi+15}.

The dissipated magnetic energy is converted into (1) kinetic energy of
the bulk outflow motion; and (2) thermal energy of the outflow. As was
long thought and recently proven numerically \citep{SS14, US14},
part of the dissipated energy is used to accelerate particles to
non-thermal distribution, rather than heat a thermal distribution of
particles to a higher temperature. In the context of energy transfer
from the magnetic field, this energy is part of the thermal energy
given to the plasma (rather than the kinetic energy). The difference
between thermal and non-thermal heating would be manifested in the rate at
which this energy could be radiated away. One generally expects that
non-thermal particles would radiatively lose their energy faster.

In the context of GRBs, if indeed cooling is efficient, (most of-) the
thermal energy will be radiated away during the prompt phase, either
as thermal photons at the photosphere, or as non-thermal photons above
it. As opposed to that, the bulk kinetic energy could not be converted
into radiation on the short time scale characterizing the prompt
phase. Instead, it will gradually dissipate during the afterglow
phase. Thus, measurements of the thermal and non-thermal energies
during the prompt phase and comparing them to the outflow kinetic
energy (that could be deduced from afterglow measurements) would put
strong constraints on the validity of the magnetized model.

In this paper, I show that the maximum ratio of thermal to kinetic
energy is in fact universal, and is independent on many of the model's
parameters. If radiative cooling is slow, the amount of energy that
can be released as thermal photons cannot exceed $(13 \hat \gamma
-14)^{-1} = 30\%$ (for adiabatic index $\hat \gamma = 4/3$) of the
kinetic energy. This energy would be released at the photosphere, and
will therefore be observed as a (modified) thermal component. I should
stress that 30\% is an absolute upper limit: since the photospheric
radius is expected to be below the saturation radius (namely, occur
while the flow still accelerates), only the thermal energy released up
until this radius could be radiated as such. One therefore expects the
observed ratio of thermal to kinetic energy to be no more than a few
\%.  On the other extreme, if radiative cooling is efficient, the
fraction of energy released as (non-thermal) photons is equal at most
to the remaining kinetic energy, regardless of the unknown model
parameters, such as the magnetization or the reconnection rate. This
implies that within the context of Poynting-flux dominated outflow,
the overall radiation observed during the prompt phase cannot exceed
the kinetic energy inferred from afterglow observations. Thus,
additional - or different - mechanisms must be operating in those GRBs
in which the energy released during the prompt phase exceeds the
kinetic energy. The obtained results are aligned, of course, with the
numerical results obtained in \citet{DS02}, however they generalize
the numerical results obtained there by providing robust,
model-independent upper limits, which can be directly compared with
observations.

This paper is organized as follows. In \S\ref{sec:2} I provide the
underlying model assumptions. I then calculate the ratio of thermal to
kinetic energy released for the slow cooling and fast cooling
scenarios in \S\ref{sec:3}. In \S\ref{sec:obs} I discuss the current
observational status of the prompt and afterglow GRB measurements, as
had been accumulated in the past decade or so. I point to gaps in the
analysis, that could lead to breakthrough in understanding the
magnetization in GRBs. I then discuss in \S\ref{sec:5} the
implications and limitations of the Poynting-flux dominated model in
view of the existing data before summarizing.

\section{Basic Model Assumptions}
\label{sec:2}

As a model of Poynting-flux dominated flow, I adopt the ``striped
wind'' model of \citet{Coroniti90}, whose dynamics were studies by
several authors \citep{Drenkhahn02, DS02, Giannios05, GS05, MR11}.
The magnetic field in the flow changes polarity on a small scale
$\lambda$ due to rotation of an inclined magnetic dipole. This scale
is of the order of the light cylinder in the central engine frame
($\lambda \approx 2\pi c/\Omega$, where $\Omega$ is the angular
frequency of the central engine - presumably a spinning black
hole). The polarity change leads to magnetic dissipation via
reconnection process, which is assumed to occur at a constant rate
along the jet. As a consequence, the magnetic field decays during a
characteristic (comoving) time scale $\tau' = \lambda'/\epsilon v_A'$,
where $v_A' \approx c$ is the comoving Alfv\'enic speed and $\lambda'
= \Gamma \lambda$, where $\Gamma$ is the Lorentz factor of the
flow. All the uncertainty in the microphysics of the reconnection
process is taken up by the dimensionless factor $\epsilon$, which is
often assumed in the literature a fixed value, $\epsilon = 0.1$.

Way above the Alfv\'enic radius (the radius in which the flow velocity
is equal to the Alfv\'en speed), the flow is assumed to be purely
radial. The dominant magnetic field component is $B = B_\phi \gg B_r,
B_\theta$. For stationary case in ideal magneto-hydrodynamics (MHD), this implies that
$\partial_r(\beta r B) = 0$, where $\beta$ is the outflow velocity and
$B$ is the magnetic field in the observer's frame. For non-ideal MHD,
the evolution of the magnetic field is given by \citep{Drenkhahn02,
  DS02} 
\beq 
\partial_r (r u b) = -{r b \over c \tau'} = -{r b \over c} {\epsilon
  \Omega \over 2 \pi \Gamma},
\label{eq:1}
\eeq
where $b = B / \sqrt{4 \pi} \Gamma$ is the (normalized) magnetic field
in the comoving frame, $u = \Gamma \beta$ and $\beta =
(1-\Gamma^{-2})^{1/2}$ is the normalized outflow velocity.

It was shown by \citet{Drenkhahn02} that for Poynting flux dominated
flow with $\Gamma \gg 1$, the flow accelerates as $\Gamma(r) \propto
r^{1/3}$. For the purpose of this work, I point out that this is a
very robust result, that is independent on the reconnection rate
($\epsilon \Omega$) and can be derived directly from Equation
\ref{eq:1}, as long as the outflow is Poynting flux dominated, $L_{pf}
= \Gamma u r^2 b^2 \gg L_k$.

This result can be understood by noting the following. First, for
$\Gamma \gg 1$, Equation \ref{eq:1} can be written as
$\partial_r(L_{pf}) \simeq \partial_r(r^2 \Gamma^2 b^2) \propto -
(rb)^2$. Using $L = L_{pf} + L_k$, conservation of energy implies
$\partial_r(L_k) = -\partial_r(L_{pf}) \propto (rb)^2$. Second, the
flux of kinetic energy can be written as $L_k = \dot M \Gamma c^2$,
where $\dot M$ is the mass ejection rate per time per sterad, which is
assumed steady. Thus, $\partial_r(L_k) \propto \partial_r
(\Gamma)$. Combined together, one obtains $\partial_r \Gamma \propto
(rb)^2$. Using now the assumption $L_{pf} \gg L_k$, one finds that $L
\simeq L_{pf} = \Gamma^2 r^2 b^2$ which implies $\Gamma^2 \partial_r
\Gamma \approx$~Const, with the solution $\Gamma(r) \propto r^{1/3}$.

{\bf Dynamic equations}. The evolution of the proper mass density,
$\rho$, energy density (excluding rest mass), $e$, the 4-velocity $u$
and the magnetic field strength, $b$ are determined by conservation of
mass, energy and momentum, together with equation \ref{eq:1}. These
are combined with the equation of state, $p = (\hat \gamma - 1) e$,
where $\hat \gamma$ is the adiabatic index. When radiative losses are
included, these equations take the form \citep{DS02}:
\beq
\partial_r(r^2 \rho u) = 0 ~~\rightarrow~~ \dot M = r^2 \rho u c,
\label{eq:2} 
\eeq
\beq
\partial_r \left[r^2 \Gamma u \left(\omega + b^2\right)\right] = -r^2 \Gamma {\Lambda \over c},
\label{eq:3}
\eeq
\beq
\partial_r \left[r^2 \left\{\left(\omega + b^2 \right) u^2 + {b^2
    \over 2} + p \right\} \right] = 2 r p - r^2 u {\Lambda \over c}.
\label{eq:4}
\eeq
Here, $\omega = e + p + \rho c^2 = \hat \gamma e + \rho c^2$ is the
proper enthalpy density, and $\Lambda$ is the (comoving) emissivity
(energy radiated per unit time per unit volume), which is assumed
isotropic in the comoving frame.

As the heated particles radiate their energy they cool. The emissivity takes the form 
\beq
\Lambda = k {e c u \over r},
\label{eq:5}
\eeq
where $k$ is an adjustable cooling length. In \citet{DS02}, a value of
$k=0$ was taken below the photosphere, justified by the fact that in
this regime the photons are coupled to the particles, while $k=10^4$
was assumed above the photosphere. This high value was justified by
the assumption of synchrotron cooling of very energetic particles in
the strong magnetic field expected in this scenario. While it is far
from being clear that the electrons can be accelerated very large
Lorentz factors in this model \citep[See][]{BPY17}, as I
show here, in fact the exact value of $k$ is of no importance, as long
as $k \gtrsim 1$.

Before solving these equations, we note that for $k=0$, the energy
equation (\ref{eq:3}) can be integrated to obtain $L = L_{pf} + L_k +
L_{th}$, where $L_k (= \dot M \Gamma c^2) = r^2 \Gamma u c \rho c^2$ is the kinetic
luminosity (per steradian) and $L_{th} = r^2 \Gamma u c \hat \gamma e$
is the thermal luminosity (per steradian). The thermal luminosity
though includes a pressure term, and therefore the available
luminosity that would be observed if the thermal energy could be
entirely released (e.g., at the photosphere) is $L_{th}^{ob.} = r^2
\Gamma u c e$. When radiative losses are included ($k >0$), one can
further define a non-thermal luminosity by $L_{NT} = L - L_{pf} -
L_{th} - L_k$.

\section{Upper limits on the ratio of radiated to kinetic energy}
\label{sec:3}

The set of Equations \ref{eq:1} -- \ref{eq:4} can be simplified for
the case $\Gamma \gg 1$, by noting that one can approximate $\Gamma u
\simeq u^2 + 1/2$. Using this in the energy Equation \ref{eq:3} and
plugging the result in the momentum Equation \ref{eq:4}, one obtains
\beq
\partial_r \left[ r^2 \left(p - {\omega \over 2} \right) \right] \simeq 2 r p + k{r e \over 2}.
\label{eq:6}
\eeq
This equation implies scaling laws on the energy and number densities,
$e= e_0 r^{-7/3}$ and $\rho = \rho_0 r^{-7/3}$.

{\bf Slow cooling scenario.} Let us first consider the case in which
$k=0$, namely radiative losses are dynamically unimportant. Using
$\omega = \hat \gamma e + \rho c^2$, $p = (\hat \gamma - 1)e$ and the
scaling laws obtained above in Equation \ref{eq:6}, this equation
becomes $\rho_0 c^2 = (13 \hat \gamma - 14) e_0$, or
\beq
{L_{th}^{ob} \over L_k} = {e_0 \over \rho_0 c^2} = {1 \over 13 \hat
  \gamma - 14} = {3 \over 10},
\label{eq:7}
\eeq
where the last equality holds for $\hat \gamma = 4/3$ (for $\hat
\gamma = 5/3$, one finds $e_0/\rho_0 c^2 = 3/23$). I stress that this
is an absolute upper limit that can be obtained only if the
photospheric radius is above the saturation radius (this depends on
the reconnection rate). Typically, this is not the case, and the ratio
$L_{th}^{ob}/L_k$ is significantly less than that obtained in equation
\ref{eq:7}.

{\bf Fast cooling scenario.} From Equation \ref{eq:6}, it is clear
that for $k \gg 4 (\hat \gamma -1) \simeq 1$, the second term in the
right hand side always dominates the first term. The scaling laws for
$e$ and $\rho$ are not changed, implying that Equation \ref{eq:6}
takes the form
\beq
\left( 1- {\hat \gamma \over 2}\right) e_0 + {\rho_0 c^2 \over 2} = {3 \over 2} k e_0,
\label{eq:8}
\eeq
or (for $k \gg 1$)
\beq
{L_{th}^{ob} \over L_k} = {e_0 \over \rho_0 c^2} \simeq  {1 \over 3k}.
\label{eq:9}
\eeq
One therefore concludes that for $k \gg 1$, the observed thermal
luminosity, $L_{th}^{ob}$ can be neglected with respect to the kinetic
luminosity, $L_k$.

Using the result $ke = \rho c^2/3$ and neglecting $L_{th}$ relative to
$L_k$ and $L_{pf}$, the energy Equation (\ref{eq:3}) takes the form
\beq
\partial_r\left(L_k + L_{pf} \right) = - {L_k \over 3 r}.
\label{eq:10}
\eeq
Since $L_{NT} \simeq L - L_{pf} - L_k$, and $L$ is constant, Equation \ref{eq:10} can be written as 
\beq
\partial_r \left( L_{NT} \right) = {L_k \over 3 r}.
\label{eq:11}
\eeq
During the acceleration episode, in the regime where $\Gamma \gg 1$
the kinetic luminosity scales as $L_k \propto r^{1/3}$ (this result is
immediately obtained from the scaling laws of $\rho$ and $\Gamma$).
Equation \ref{eq:11} therefore implies both a similar scaling law of
$L_{NT} \propto r^{1/3}$, and a similar scaling coefficient. This
means that at the end of the acceleration, $L_{NT} = L_K$, namely up
to 1/2 of the dissipated magnetic energy could be radiated away, while
the other 1/2 remains in the form of kinetic energy. Further note that
this result is very robust, and is independent on any of the unknown
model parameters, neither on the adiabatic index. It holds for any
value of $k \gg 1$. Similar to the thermal emission calculation, this
is an upper limit, which depends on the assumption of strong
emissivity along the jet. In reality, all emission mechanism
(synchrotron, Compton, Bremsstrahlung, etc.) will decay with radius,
making the observed non-thermal energy to be less than the kinetic
energy \citep{BP15, BPY17}

In order to demonstrate the validity of the analytical approximations
used in deriving these conclusions, I have solved numerically the
exact set of Equations \ref{eq:1} -- \ref{eq:4}, to find the radial
evolution of the dynamical variables ($\Gamma$, $e$, $\rho$ and $b$)
and the derived variables (such as $L_{pf}$, $L_k$, $L_{th}$, $L_{NT}$
and $u$). These set of equations are coupled and stiff, and thus in
order to solve them, I first re-wrote them in terms of a variable
$\vec A = \left\{ L_{pf},~ L_k,~L_{th},~u \right\}$, and then
calculated $d\log(\vec A)/d\log(r)$. When formulated in this way,
standard numerical ordinary differential equation (ODE) solver could
be used.

\begin{figure}
\plotone{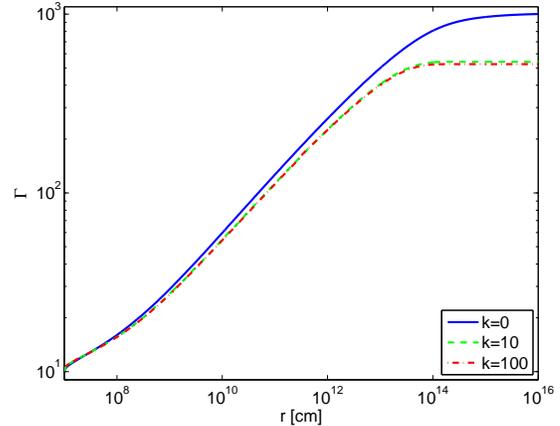}
\caption{Evolution of the bulk Lorentz factor}
\label{fig:1}
\end{figure}

\begin{figure}
\plotone{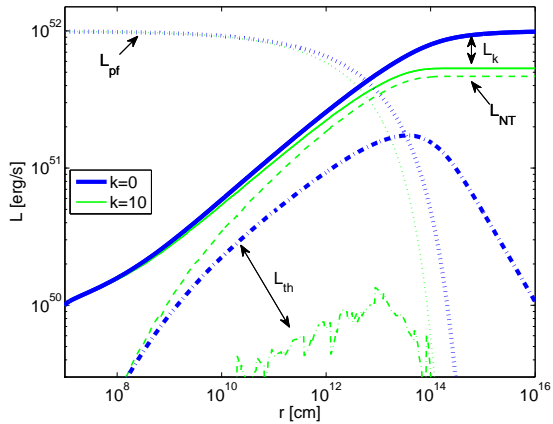}
\caption{Radial evolution of the luminosities. Think (blue) curves are
  for $k=0$ scenario, while thin (green) are for the radiative case
  with $k=10$. Solid: $L_k$, dotted: $L_{pf}$, dash-dotted: $L_{th}$
  and dash: $L_{NT}$. As explain in the text, for large $k$, the
  magnetic energy is equally distributed between kinetic and radiated
  energy, and thus $L_{NT}$ approaches $L_k$.}
\label{fig:2}
\end{figure}

\begin{figure}
\plotone{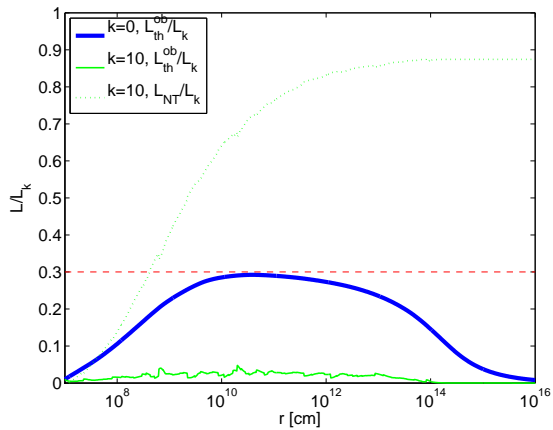}
\caption{Ratio of observed to kinetic luminosities. Thick (blue):
  $k=0$ scenario. $L_{th}^{ob}/L_k = 0.3$ (for adiabatic index $\hat
  \gamma = 4/3$). Thin (green): $k=10$. The non-thermal radiated
  energy reaches slightly less than 90\% of the kinetic energy, while
  the thermal energy is negligible in this scenario.}
\label{fig:3}
\end{figure}

The results of the numerical calculations are shown in Figures
\ref{fig:1} -- \ref{fig:3}. In producing the results, I chose as initial
conditions $L = 10^{52}~\rm{erg~s^{-1}~sterad^{-1}}$, initial
magnetization parameter $\sigma_0 \equiv L_{pf,0}/L_{k,0} = 100$ at
$r_0 = 10^7$~cm (corresponding to an initial 4-velocity $u_0 =
\sqrt{\sigma_0} = 10$), and reconnection rate $\epsilon\Omega =
10^3~{\rm s^{-1}}$. The flow was assumed initially cold ($e_0 \equiv
e|_{r_0} = 0$), and adiabatic index $\hat \gamma = 4/3$ assumed (this
is relevant below the photosphere). I chose three different values of
$k = 0, 10, 100$ representing possible different emissivities.

In Figure \ref{fig:1} I present the evolution of the bulk Lorentz
factor. For $k=0$ case, the outflow terminates at $\Gamma \simeq
\sigma_0^{3/2} = 1000$, as predicted by \citet{Drenkhahn02}. In the
radiative scenarios, the terminal Lorentz factor is slightly above
half of that value (540 and 525 for the $k=10, 100$ scenarios,
respectively), in accordance with the finding that slightly over half of
the final energy is in kinetic form. 

In figure \ref{fig:2} I show the radial evolution of the various
luminosities: $L_{pf}$, $L_k$, $L_{th}$ and $L_{NT}$ for the $k=0$ and
$k=10$ scenarios (the results obtained for $k=100$ are very similar to
the ones obtained for $k=10$, and are thus omitted for clarity). All
the numerical results are in accordance with the analytical
calculations presented above. In particular, when non-thermal
radiation is omitted ($k=0$), the ratio between $L_{th}$ and $L_k$
approaches 30\%. This is directly seen in Figure \ref{fig:3}. In the
radiative scenario ($k=10$), this ratio is much lower; on the other
hand, $L_{NT}$ approaches $L_k$, as is seen in Figures \ref{fig:2} and
\ref{fig:3}.

\section{Observational constraints}
\label{sec:obs}

The results derived above provide two clear predictions about the
upper limits of thermal and non-thermal luminosities that can be
expected during the prompt emission phase in GRBs. These can be tested with current and
future observations, that can therefore be used to test the validity
of this model. Identification and analysis of the properties of a
thermal component in GRBs is a relatively new field, and, as a result,
only sparse data exists to date (see further discussion below). On the
other hand, calorimetry of the total radiative efficiency, namely the
non-thermal emission of the prompt phase has been carried out
extensively in the past two decades, since the discovery of GRB
afterglow. It is therefore useful in the context of this work to
briefly summarize the existing observational status.

It is common to define the efficiency of the prompt emission radiation
as the ratio of the total energy released in gamma rays (thermal plus
non-thermal), divided by the the total (radiative plus kinetic)
energy,
\beq
\eta \equiv {E_\gamma \over (E_k + E_\gamma).}
\label{eq:12}
\eeq
Within the limits of the observed spectral band, the GRB prompt
emission provides a direct probe of the energy released during the
prompt phase. The kinetic energy, on the other hand, is estimated by
fitting the afterglow observations, which, for historical reasons, are
typically available at 11 hours in the x-ray band. The fitting is done
within the framework of the ``classical'' synchrotron model, under the
assumption that electrons accelerated to a power law distribution in
the propagating forward shock wave. The advantage is that at this
time, the reverse shock should already disappear, and the temporal and
spectral evolution of the emitted signal should be well characterized
by simple scaling laws \citep{BM76, MLR93, MR97, SPN98, vPKW00, GS02},
thereby enabling a reasonably accurate estimate of the outflow kinetic
energy.

Few early works that estimated the efficiency of various samples of
bursts were carried by \citet{Kumar00, FW01, PK02, Yost+03, LZ04,
  Zhang+07b, Racusin+11, DAvanzo+12}. More recent works not only
estimated by efficiency but further the absolute released energy by
correcting for the finite jet opening angle \citep{Cenko+10, Cenko+11,
  Troja+12, Guidorzi+14, Laskar+14, Laskar+15, Laskar+16}. The
different samples consist of bursts observed by different instruments
thereby having different spectral coverage, and used several different
methods in estimating the efficiencies. It is therefore impossible to
combine all the collected data into one big sample.

Despite these differences, it is very interesting that all these works
arrive at the same conclusion, namely that the efficiency varies
widely between different bursts within the same sample. It is consistently
found that the efficiency is ranging from less than 1\% to over 90\%,
with about 1/2 of GRBs in each sample showing efficiency of over
50\%. Note that in the notation used in this paper, radiative
efficiency of $\eta > 50\%$ is equivalent to $L_{NT} > L_k$, and is
thus forbidden within the framework of the Poynting-flux dominated model. As a
concrete example, in the analysis carried by \citet{Laskar+15}, 13/24
GRBs show $E_\gamma > E_k$, with 11/24 being more than 1~$\sigma$ away
from $E_\gamma \leq E_k$. This result is not unique, but is rather
representative of all other analyses mentioned above. Interestingly, a
similar conclusion was reached when analyzing a large sample of short
GRBs \citep{Fong+15}, implying that the large range of radiative
efficiencies appear in both the long and short GRB populations.

This large range of efficiencies found is challanging to all
theoretical models, that need to explain both the very high efficiency
seen in tens of \% of the GRB population, as well as the wide
separation between the bursts. Motivated by this cahllange, a
different analysis method was proposed by \citet{Beniamini+15,
  Beniamini+16}. In these works, the authors argue that the late time
x-ray flux may not be a good proxy to the kinetic energy, due to
either significant synchrotron self Compton radiation that lowers the
synchrotron flux at the observed band, or alternatively a weaker than
expected magnetic field that prevents an efficient radiation at the
x-ray band (``slow cooling''). Instead, they propose to use the GeV
emission seen at much earlier times (hundreds of seconds, though later
than the observed time of photons at lower energies) in several
LAT-detected GRBs as a better proxy to the remaining kinetic energy.

Using the GeV photons to estimate the kinetic energy results in
significantly higher kinetic energy than that inferred from the x-rays
and corresponding lower efficiency, which is found to be around 15\%.
While this model seem to overcome the high efficiency problem, the use
of the GeV emission as a proxy to the kinetic energy needs to be done
with great care. First, the inferred energy is much higher (up to two
orders of magnitude) than the typical energies inferred at very late
times \citep{SB11}. Second, it is far from being clear that the GeV
photons have synchrotron (rather than inverse Compton) origin, as is
assumed in these fits. Third, it is not obvious that the GeV photons
originate from the forward shock. On the contrary, it was shown by
\citet{PW05b} that a significant fraction of the GeV photons expected
at this time of hundreds of seconds originate from the reverse shock
that could well exist at this epoch, and are upscattered by electrons
at the forward shock. Finally, a detailed analysis (Gomptz et. al., in
prep.) shows that about half of the LAT bursts in the sample used by
\citet{Beniamini+15} are in fact in the fast cooling regime.

Nonetheless, the high eficiency inferred by many authors is a major
challange not only to the magnetized model presented here, but to the
alternative ``internal shocks'' model as well. There is a broad
agreement that the radiative efficiency expected in the internal shock
model is not likely to exceed a few \% - few tens of \% as most
\citep{Kobayashi+97, DM98, GSW01, Ioka+06, Peer+17}, though it was
shown that under extreme conditions, higher efficiency could be
reached in this model \citep[e.g.,][]{Beloborodov00, KS01}.

There are two important possible caveats of the efficiency estimates
presented in the literature to date, which are related to estimate of
the kinetic energy, $E_k$ from afterglow data. The first is that many
of the above mentioned works assume a ``top hat'' jet, namely neglect
any internal jet structure of the form $\Gamma = \Gamma (\theta)$
($\Gamma$ is the jet Lorentz factor and the angle $\theta$ is measured
from the jet axis). On the other hand, simple phase space argument
implies that in most observed GRBs the observer is located off the jet
axis. While the structure of the jets are unknown, angle-dependent
Lorentz factor could lead to an uncertainty in the estimated value of
the kinetic energy by a factor of up to a few.

A second caveat relates to the microphysics of particle
acceleration. Despite major progress in recent years in understanding
particle acceleration in shock waves, the fraction of electrons
accelerated to high energies in relativistic shocks is still
uncertain. As was pointed out by \citet{EW05}, there is a degeneracy
in interpreting the observed afterglow signal between the fraction of
particles accelerated (and emitting the radiation), the jet kinetic energy
as well as other parameters such as the density and the magnetic
field. Thus, it is possible to explain the observed signal in a model
in which a relatively small fraction of electrons are accelerated,
provided that the kinetic energy is high, which will reflect in a
lower efficiency than estimated by the works discussed above.

While these caveats put strong constraints on the ability of current
models to accurately estimate the prompt phase efficiency, in recent
years there is a major progress in modeling structured jets observed
off-axis \citep[e.g.,][]{ZM09, VEM12, VEM13}, which enable to break
some of the degeneracy involved in the jet structure and viewing angle
\citep{Ryan+15a}. Similarly, advance in numerical,
particle-in-cell (PIC) simulations enabled much better understanding
of the microphysics of particle acceleration \citep{Spit08, SS11},
which could be used to constrain the fractions of non-thermal and
thermal particles heated by the external shock \citep{GS09}. It is
thus anticipated that much better observational constraints on the
efficiency will become available in the near future.

A second prediction of the magnetized model presented here is an upper
limit on the ratio of the released thermal energy to kinetic
energy. In recent years it became evident that thermal emission
component does exist in several bright GRBs, such as GRB090902B
\citep{Abdo+09a, Ryde+11}, GRB100724B\citep{Guiriec+11},
GRB090510\citep{Ackermann+10}, GRB110721A \citep{Axelsson+12,
  Iyyani+13}, GRB110920A \citep{Iyyani+15} and others. In a few bright
GRBs, such as GRB090902B, it clearly dominated the spectrum.

However, no systematic analysis was carried so far about the relative
strength and ubiquitousness of the thermal component. Partially, this
is due to the fact that a firm detection of a thermal component is
relatively difficult, as (1) it requires a different template than the
commonly used ``Band'' function fit, and (2) there are various efects
that act to smear (broaden) the signal; see \citet{PR17} for
details. Despite this fact, there is an increasing evidence that a
thermal component is more ubiquioutous among bright GRBs (Ryde
et. al., in prep.). This can be understood, as a clear detection of a
thermal component requires a more refined template in fitting the
observed spectrum. It is therefore more difficult to detect in weak
GRBs, in which the number of photons are limited. If confirmed, this
result therefore suggests that a thermal component might in fact be
very ubiquitous; in many GRBs in which it is not detected, it is
mainly due to technical reasons, as, due to their cosmological
  distribution and the detectors limitations, most GRBs are detected
close to the detector's limit.  The results presented here thus raise
the need for a more comprehensive analysis of both the thermal and the
non-thermal flux, as a key way in constraining the outflow
magnetization.

\section{Implications and limitations}
\label{sec:5}

The energy released as thermal and non-thermal during the acceleration
phase ($L_{th}$ and $L_{NT}$) would be directly observed during the
prompt phase, either as thermal photons released at the photosphere,
or as non-thermal photons released at larger radii. As opposed to
that, the kinetic energy ($L_k$) could not be directly observed during
the prompt phase, unless an additional, non-magnetized dissipation
process takes place. Such process might be shock waves, that could
develop as a result of instabilities in the flow. In particular, these
might be expected at large radii, where the magnetization is weak; we
discuss this phenomenon below. However, as pointed in the literature,
substantial kinetic energy released by collisions is generally less
favourable, due to the low efficiency of this process.

Instead, the kinetic energy will be gradually released during the
afterglow phase: it will be used to accelerate and heat particles from
the ambient medium, which will radiate the observed afterglow. As
discussed in \S\ref{sec:obs} above, the very high efficiency of
radiation during the prompt phase, if indeed confirmed, challanges the
validity of the magnetized model.

This difficulty adds to the difficulty of magnetized models to account
for a significant, sub-MeV thermal component \citep{ZP09, BP15} as is
reported in several bursts.  Interestingly, very similar results are
obtained within the reconnection model suggested by \citet{LK01}, in
which different assumptions about the reconnection rate lead to
different scaling law $\Gamma \propto r^{1/2}$.

One possible solution within the framework of Poynting-flux dominated
flows is to invoke a more complicated dissipation scheme. For example,
one may assume that the dissipation of the magnetic energy does not
occur continuously along the jet, but only in specific regions. This
could be triggered, e.g., by outflow discontinuities such as
turbulence \citep[e.g.,][]{ZY11}. However, detailed numerical models
carried so far of more complicated outflow dynamics \citep{
  Deng+15} did not reveal a substantially different dynamics than the
simple 1-d model analyzed in this work, nor better efficiency than
derived here.  Another possibility is Compton drag (namely, the emitted
radiation is non-isotropic in the comoving frame). In this scenario,
somewhat higher efficiency could be achieved under the appropriate
conditions \citep{LG16}.

A detailed model by \citet{MU12} suggested that the rate of
reconnection changes along the the jet, from being slow (Sweet-Parker
like geometry) at small radii to fast (Petschek-like geometry) at
larger radii. This transition is initiated by a change in the plasma
conditions, from being collisional to collisionless. The transition
could occur if certain conditions are met, such as the production of a
large number of pairs in the inner jet regions which later annihilate
at large radii. Despite the different underlying assumptions,
\citet{MU12} find that the outflow dynamics in this scenario is not
substantially different than the one considered here. The main
application of this scenario would therefore be a reduction of the
relative strength of the thermal component, as most of the dissipation
occurs above the photosphere.

The simplified dynamics considered here may be modified by variations
in the conditions at the base of the jet. Such fluctuations could lead
to internal shocks, which would further dissipate part of the kinetic
energy, thereby increase the efficiency of the prompt emission beyond
the values calculated here. This is due to the fact that additional
source of energy (associated with the relative motion of the outflow)
is added to the energy associated with the dissipation of the magnetic
field.

In general, internal shocks are expected in region of low
magnetization, namely $\sigma < 1$. While the focus in this work is on
highly magnetized flows, as shown in Section \ref{sec:3} above due to
the disipation of the magnetic energy that is used to accelerate the
flow, at sufficiently large radii the magnetic energy becomes
sub-dominant (see Figure \ref{fig:2}). Furthermore, in a scenario of
variable, magnetized outflow, magnetic energy conversion inside a
plasma shell may occur directly as a result of magnetic pressure
within a shell \citep{Granot+11}. These internal shocks can dissipate
a substantial fraction of the differential kinetic energy between the
shells, in both the low-magnetic as well as magnetized shell scenarios
\citep{Granot12}. 

A detailed calculation of the modification of the efficiency
calculated here due to internal shocks is beyond the scope of this
work. This is due to the fact that in this scenario, the efficiency of
kinetic energy conversion depends on the configuration of the magnetic
fields, as well as the initial conditions at the jet base.  Several
works that dealt with strong toroidal field concluded that the
efficiency is not expected to be high, typically a few \% at most for
magnetization parameter $\sigma \geq 10$ \citep{KC84, ZK05,
  Narayan+11, Komissarov12}. It is found in these works that there is
an inverse correlation: a stronger magnetization leads to a lower
efficiency in energy conversion by shock waves.  

Furthermore, if the magnetic field have a strong poloidal component,
the formation of shock waves is suppressed, and it is not clear that
the shock waves could be formed at all \citep{Bret+17}. These results
therefore suggest that the efficiency dervied here might not be
heavily modified in the present of internal shocks, if they occur
  in a regime dominated by poloidal field.

Alternatively, most of the prompt emission photons may have a
different origin. An appealing alternative is emission from the
photosphere. Thermal photons may exist in the plasma at early stages,
and advect with it until the photosphere, in which they
decouple. Thus, their existence does not require high efficiency in
kinetic or magnetic energy conversion. Indeed, there are increasing
evidence that a significant thermal component is ubiquitous
\citep[e.g.,][and discussion in \S\ref{sec:obs} above]{Lazzati+13}.
In recent years, it was demonstrated that the observed spectrum of
photons emerging from the photosphere can deviate substantially from
the naively expected ``Planck'' spectrum. This is due to both light
aberration effects \citep{Peer08, LPR13} as well as possible
sub-photospheric energy dissipation \citep[e.g.,][]{PMR06}.

\section{Summary and conclusions}
\label{sec:summary}

In this paper, I consider the ``striped wind'' model of a
Poynting-flux dominated outflow in GRBs, in which the main source of
energy is dissipation of magnetic fields. The dissipated magnetic
energy is used to both accelerate the outflow and heat particles in
the plasma, which then radiate, producing both thermal and non-thermal
emission.  I derived here simple analytical upper limits on the ratio
of thermal to kinetic energy, $L_{th}/L_k = 30\%$ (Equation
\ref{eq:7}) and non-thermal to kinetic energy $L_{NT}/L_k = 50\%$
(Equation \ref{eq:11}), and confronted these upper limits with
observations in \S\ref{sec:obs}.

The analytical upper limits and numerical results derived here are
aligned with the numerical results obtained by \citet{DS02}, albeit a
larger value of $k$ (very fast cooling) was used in that work.  The
ratio of thermal to total kinetic energy was calculated previously
only numerically, by \citet{DS02}, and later on by \citet{GS07}. These
works found that this ratio is at the range of 20\% - 30\% (though a
maximum value of 35\% was found numerically) for the parameter range
considered. The analytical results derived here thus provide a simple
explanation to the numerical works. Similarly, the ratio of 50\% of
non-thermal to kinetic energy was derived numerically by \citet{DS02}.
A heuristic, yet insightful argument for the validity of this result
was provided by \citet{SD04}\footnote{The argument appears in full
  only in the arXiv version of the proceedings.}.  The present work
thus generalizes previous treatments of the striped-wind model
scenario, by providing analytical arguments that prove the robustness
of the upper limits obtained on the ratios of both the thermal and
non-thermal fluxes. These limits are independent on uncertainties such as the
initial magnetization parameter, magnetic dissipation rate, cooling
rate or adiabatic index.

By now there are ample of works measuring the radiative efficiency of
non-thermal GRB prompt emission.  Despite the use of different samples
and different methods, a repeated result is that the efficiency
considerably varies among GRBs within the same sample, with about half
the bursts showing efficiency greater than the allowed by the
magnetized outflow scenario. As discussed in Section
  \ref{sec:obs} above, there is a large uncertainty in the estimate of
  the efficiency due to the unknown jet structure and the microphysics
  of particle acceleration. This high efficiency is further difficult
to explain within the framework of the alternative ``internal shocks''
model as well. It thus calls for a deep re-analysis using both broad
band and time dependent data, to validate these results.

A key result of this work is the upper limit on the thermal
flux. While it is clear today that a thermal emission component exists
in many GRBs, it is still uncertain how ubiquitous it is, and how
strong it is among different bursts. The analysis carried here thus
calls for a reanalysis of GRB prompt emission in order to identify
thermal compoent that could constrain the magnetization. Indeed, as
was already pointed out, e.g., by \citet{ZP09} and is further
strengthen here, identification of a strong thermal component is likely
the best observational way of constraining the outflow magnetization.

It is clear that the dynamical model used here is simplified, as it
cannot account for variation in the outflow. However, the rate of
reconnection depends on the exact configuration of the magnetic field,
and therefore can only be tracked by numerical MHD models. Existing
numerical results \citep{MU12} suggest that the outflow dynamics may
in fact be close to the simplified model used here. This fact may
further be used to constrain the validity of the magnetized model in
explaining the dynamics of GRB outflows, in particularly in those GRBs
in which the lightcurve is highly variable. It may further suggest a
correlation between the observed lightcurve variability and the
existence of a strong thermal component, as both are characterizing
outflows which are only weakly magnetized.

\acknowledgments AP wishes to thank Damien Begue, Brad Cenko, Amir
Levinson, Yuri Lyubarski, Raffaella Margutti and Bing Zhang for useful
comments. I further wish to thank the anonymous referee for useful
comments that helped improving this manuscript. This research was
partially supported by the European Union Seventh Framework Programme
(FP7/2007-2013) under grant agreement ${\rm n}^\circ$ 618499. I
further wish to acknowledge support from NASA under grant
\#NNX12AO83G.

\bibliographystyle{/Users/apeer/Documents/Bib/apj}



\end{document}